\documentclass[11pt,twoside]{article}
\usepackage{graphicx}
\usepackage{amsmath}
\usepackage{amssymb}
\usepackage{cite}

\begin{document}
\newcommand{\pst}{\hspace*{1.5em}}

\newcommand{\rigmark}{\em Journal of Russian Laser Research}
\newcommand{\lemark}{\em Volume 30, Number 5, 2009}

\newcommand{\be}{\begin{equation}}
\newcommand{\ee}{\end{equation}}
\newcommand{\bm}{\boldmath}
\newcommand{\ds}{\displaystyle}
\newcommand{\bea}{\begin{eqnarray}}
\newcommand{\eea}{\end{eqnarray}}
\newcommand{\ba}{\begin{array}}
\newcommand{\ea}{\end{array}}
\newcommand{\arcsinh}{\mathop{\rm arcsinh}\nolimits}
\newcommand{\arctanh}{\mathop{\rm arctanh}\nolimits}
\newcommand{\bc}{\begin{center}}
\newcommand{\ec}{\end{center}}

\thispagestyle{plain}

\label{sh}


\begin{center} {\Large \bf
\begin{tabular}{c}
ENTANGLEMENT AND MIXEDNESS
\\[-1mm]
IN OPEN SYSTEMS WITH CONTINUOUS VARIABLES
\end{tabular}
 } \end{center}

\bigskip

\bigskip

\begin{center} {\bf

Aurelian Isar
}\end{center}

\medskip

\begin{center}
{\it
National Institute of Physics and Nuclear Engineering,\\
P.O.Box MG-6, Bucharest-Magurele, Romania\\
}
\smallskip

$^*$Corresponding author e-mail:~~~isar@theory.nipne.ro\\
\end{center}

\begin{abstract}\noindent

In the framework of the theory of open systems based on completely positive quantum dynamical semigroups, we give a description of the dynamics of entanglement for a system consisting of two uncoupled modes interacting with a thermal environment. Using Peres-Simon necessary and sufficient criterion of separability for two-mode Gaussian states, we describe the evolution of entanglement in terms
of the covariance matrix for a Gaussian input state.  We determine the asymptotic Gaussian maximally entangled mixed states (GMEMS) and their corresponding asymptotic maximal logarithmic negativity, which characterizes the degree of entanglement.
Using the symplectic
eigenvalues of the asymptotic covariance matrix, the expressions of von Neumann entropy and mutual information
of asymptotic GMEMS are obtained.

\end{abstract}

\medskip

\section{Introduction}
\pst

Quantum entanglement of Gaussian states constitutes a fundamental
resource in continuous variable (CV) quantum information processing and communication \cite{bra1}. For the class of Gaussian states there exist necessary and sufficient criteria of entanglement \cite{sim,dua} and quantitative entanglement measures \cite{vid,gie}.The quantum information processing tasks are difficult to implement, due to the fact that any realistic quantum system is not isolated and it always interacts with its environment. Quantum coherence and entanglement of quantum systems are inevitably influenced during their interaction with the external environment. As a result of the irreversible and uncontrollable phenomenon of quantum decoherence, the purity and entanglement of quantum states are in most cases degraded. Therefore in order to describe realistically quantum information processes it is necessary to take decoherence and dissipation into consideration. Decoherence and dynamics of quantum entanglement in CV open systems have been intensively studied in the last years \cite{oli,pra,ser1,ben1,avd,man,aphysa,aeur}.

When two systems are immersed in an environment, then, in addition to and at the same time with the quantum decoherence phenomenon, the environment can also generate a quantum entanglement of the two systems and therefore an additional mechanism to correlate them \cite{ben1,vvd1,ben2}. In this paper we study, in the framework of the theory of open systems based on completely positive quantum dynamical semigroups, the dynamics of the CV entanglement of two modes (two identical harmonic oscillators) coupled to a common thermal environment.  We are interested in discussing the correlation effect of the environment, therefore we assume that the two modes are uncoupled, i.e. they do not interact directly. The initial state of the subsystem is taken of Gaussian form and the evolution under the quantum dynamical semigroup assures the preservation in time of the Gaussian form of the state. In Sec. 2 we write the Markovian master equation in the Heisenberg representation for two uncoupled harmonic oscillators interacting with a general environment and the evolution equation for the covariance matrix of the considered system. By using the Peres-Simon criterion of separability for two-mode Gaussian states \cite{sim,per}, we investigate in Sec. 3 the dynamics of entanglement for the considered system. We determine the asymptotic Gaussian maximally entangled mixed states (GMEMS) and their corresponding asymptotic maximal logarithmic negativity, which characterizes the degree of entanglement. During the interaction with the environment,
any pure quantum state evolves into a mixed state and for quantum information processes
it is essential to determine the degree of mixedness of quantum
states, which can be characterized
either by the von Neumann entropy
or by the linear entropy \cite{ser2}. Using the symplectic
eigenvalues of the asymptotic covariance matrix, we obtain the expressions of von Neumann entropy and mutual information
of asymptotic GMEMS, which
quantifies the total amount of correlations
(both quantum and classical) contained in these states.
A summary is given in section 4.

\section{Time evolution of covariance matrix for two harmonic oscillators}
\pst

We study the dynamics of the subsystem composed of two identical non-interacting oscillators in weak interaction with a thermal environment. In the axiomatic formalism based on completely positive quantum dynamical semigroups, the irreversible time evolution of an open system is described by the following general quantum Markovian master equation for an operator $A$ in the Heisenberg representation ($\dagger$ denotes Hermitian conjugation) \cite{lin,rev}:
\begin{eqnarray}\frac{dA(t)}{dt}=\frac{i}{\hbar}[H,A(t)]+\frac{1}{2\hbar}\sum_j(2V_j^{\dagger}AV_j - V_j^{\dagger}
V_jA-AV_j^{\dagger}V_j).\label{masteq}\end{eqnarray}
Here, $H$ denotes the Hamiltonian of the open system and the operators $V_j, V_j^\dagger,$ defined on the Hilbert space of $H,$ represent the interaction of the open system with the environment.

We are interested in the set of Gaussian states, therefore we introduce such quantum dynamical semigroups that preserve this set during time evolution of the system. Consequently $H$ is taken to be a polynomial of second degree in the coordinates $x,y$ and momenta $p_x,p_y$ of the oscillators and $V_j,V_j^{\dagger}$ are taken polynomials of first degree in these canonical observables. Then in the linear space spanned by coordinates and momenta there exist only four linearly independent operators $V_{j=1,2,3,4}$ \cite{san}: \begin{eqnarray}
V_j=a_{xj}p_x+a_{yj}p_y+b_{xj}x+b_{yj}y,\end{eqnarray} where
$a_{xj},a_{yj},b_{xj},b_{yj}$ are complex coefficients. The Hamiltonian $H$ of the two uncoupled identical harmonic oscillators of mass $m$ and frequency $\omega$ is given by \begin{eqnarray}
H=\frac{1}{2m}(p_x^2+p_y^2)+\frac{m\omega^2}{2}(x^2+y^2).\end{eqnarray}

The evolution given by a dynamical semigroup implies the positivity of the matrix formed by the scalar products of the four vectors
${\bf a}_x, {\bf b}_x,{\bf a}_y, {\bf b}_y$ whose entries are the components $a_{xj},b_{xj},a_{yj},b_{yj},$ respectively. We take this matrix of the following form, where all diffusion coefficients $D_{xx}, D_{xp_x},$... and dissipation constant $\lambda$ are real quantities (we put from now on $\hbar=1$):
\begin{eqnarray} \left(\begin{matrix}D_{xx}&- D_{xp_x} - i \lambda/2&D_{xy}& -
D_{xp_y} \cr - D_{xp_x} + i \lambda/2&D_{p_x p_x}&-
D_{yp_x}&D_{p_x p_y} \cr D_{xy}&- D_{y p_x}&D_{yy}&- D_{y p_y}
- i \lambda/2 \cr - D_{xp_y} &D_{p_x p_y}&- D_{yp_y} + i
\lambda/2&D_{p_y p_y}\end{matrix}\right).\label{coef} \end{eqnarray}
It follows that the principal minors of this matrix are positive or zero. From
the Cauchy-Schwarz inequality the following relations hold for the coefficients defined in Eq. (\ref{coef}): \begin{eqnarray}
D_{xx}D_{yy}-D^2_{xy}\ge0,~~~D_{xx}D_{p_xp_x}-D^2_{xp_x}\ge\frac{\lambda^2}{4},~~~
D_{xx}D_{p_yp_y}-D^2_{xp_y}\ge 0, \nonumber\\ D_{p_xp_x}D_{p_yp_y}-D^2_{p_xp_y}\ge 0,~~~ D_{yy}D_{p_yp_y}-D^2_{yp_y}\ge\frac{\lambda^2}{4},~~~
D_{yy}D_{p_xp_x}-D^2_{yp_x}\ge 0.
\label{coefineq}\end{eqnarray}

We introduce the following $4\times 4$ bimodal covariance matrix:
\begin{eqnarray}\sigma(t)=\left(\begin{matrix}\sigma_{xx}(t)&\sigma_{xp_x}(t) &\sigma_{xy}(t)&
\sigma_{xp_y}(t)\cr \sigma_{xp_x}(t)&\sigma_{p_xp_x}(t)&\sigma_{yp_x}(t)
&\sigma_{p_xp_y}(t)\cr \sigma_{xy}(t)&\sigma_{yp_x}(t)&\sigma_{yy}(t)
&\sigma_{yp_y}(t)\cr \sigma_{xp_y}(t)&\sigma_{p_xp_y}(t)&\sigma_{yp_y}(t)
&\sigma_{p_yp_y}(t)\end{matrix}\right).\label{covar} \end{eqnarray}
The problem of solving the master equation for the operators in Heisenberg representation can be transformed into a problem of solving first-order in time, coupled linear differential equations for the covariance matrix elements. Namely, from Eq. (\ref{masteq}) we obtain the following system of equations for the quantum correlations of the canonical observables, written in matrix form \cite{san}:
\begin{eqnarray}\frac{d \sigma(t)}{dt} = Y \sigma(t) + \sigma(t) Y^{\rm T}+2 D,\label{vareq}\end{eqnarray} where
\begin{eqnarray} Y=\left(\begin{matrix} -\lambda&1/m&0 &0\cr -m\omega^2&-\lambda&0&
0\cr 0&0&-\lambda&1/m \cr 0&0&-m\omega^2&-\lambda\end{matrix}\right),~~~
D=\left(\begin{matrix}
D_{xx}& D_{xp_x} &D_{xy}& D_{xp_y} \cr D_{xp_x}&D_{p_x p_x}&
D_{yp_x}&D_{p_x p_y} \cr D_{xy}& D_{y p_x}&D_{yy}& D_{y p_y}
\cr D_{xp_y} &D_{p_x p_y}& D_{yp_y} &D_{p_y p_y} \end{matrix}\right).\end{eqnarray}
The time-dependent
solution of Eq. (\ref{vareq}) is given by \cite{san}
\begin{eqnarray}\sigma(t)= M(t)[\sigma(0)-\sigma(\infty)] M^{\rm
T}(t)+\sigma(\infty),\label{covart}\end{eqnarray} where the matrix $M(t)=\exp(Yt)$ has to fulfill
the condition $\lim_{t\to\infty} M(t) = 0.$
In order that this limit exists, $Y$ must only have eigenvalues
with negative real parts. The values at infinity are obtained
from the equation \begin{eqnarray}
Y\sigma(\infty)+\sigma(\infty) Y^{\rm T}=-2 D.\label{covarinf}\end{eqnarray}

\section{Dynamics of continuous variable entanglement}
\pst

\subsection{Time evolution of entanglement}

The characterization of the separability of CV states using second-order moments of quadrature operators was given in Refs. \cite{sim,dua}. A two-mode Gaussian state is separable if and only if the partial transpose of its density matrix is non-negative [necessary and sufficient positive partial transpose (PPT) criterion]. A two-mode Gaussian state is entirely specified by its covariance matrix (\ref{covar}), which is a real, symmetric and positive matrix with the block structure
\begin{eqnarray}
\sigma(t)=\left(\begin{array}{cc}A&C\\
C^{\rm T}&B \end{array}\right),\label{cm}
\end{eqnarray}
where $A$, $B$ and $C$ are $2\times 2$ Hermitian matrices. $A$ and $B$ denote the symmetric covariance matrices for the individual one-mode states, while the matrix $C$ contains the cross-correlations between modes. Simon \cite{sim} derived the following PPT criterion for bipartite Gaussian
CV states: the necessary and sufficient criterion of separability is
$S(t)\ge 0,$ where \begin{eqnarray} S(t)\equiv\det A \det B+\left(\frac{1}{4} -|\det
C|\right)^2- {\rm Tr}[AJCJBJC^{\rm T}J]- \frac{1}{4}(\det A+\det B)
\label{sim1}\end{eqnarray} and $J$ is the $2\times 2$ symplectic matrix
\begin{eqnarray}
J=\left(\begin{array}{cc}0&1\\
-1&0\end{array}\right).
\end{eqnarray}
Since the two oscillators are identical, it is natural to consider environments for which $D_{xx}=D_{yy},~ D_{xp_x}=D_{yp_y},~D_{p_xp_x}=D_{p_yp_y},~ D_{xp_y}=D_{yp_x}.$ Then both unimodal covariance matrices are equal, $A=B,$ and the entanglement matrix $C$ is symmetric.

In order to describe the dynamics of entanglement, we used the PPT criterion \cite{sim,per} according to which a state is entangled if and only if the operation of partial transposition does not preserve its positivity. We analyzed the time evolution of the Simon function $S(t)$ (\ref{sim1}) for a thermal environment characterized by the temperature $T$ and the diffusion coefficients
 \begin{eqnarray}m\omega D_{xx}=\frac{D_{p_xp_x}}{m\omega}=
\frac{\lambda}{2}\coth\frac{\omega}{2kT},~~~D_{xp_x}=0,
~~~m^2\omega^2D_{xy}=D_{p_xp_y},\label{envcoe}\end{eqnarray} corresponding to the case when the asymptotic state is a Gibbs state \cite{rev}.
We described the generation and evolution of entanglement in terms
of the covariance matrix for a Gaussian input state and showed that for some values of the temperature of environment, the state keeps for all times its initial type:
separable or entangled. Depending on the values of temperature, mixed diffusion coefficients and dissipation constant, entanglement generation, entanglement sudden death or a repeated collapse and revival of entanglement can take place \cite{arus1}.

\subsection{Logarithmic negativity}

In order to quantify the degree of entanglement of the infinite-dimensional system states of the two oscillators it is suitable to use the logarithmic negativity. For Gaussian states, the measures of entanglement of bipartite systems are based on some invariants constructed from the elements of the covariance matrix \cite{oli,avd}. Heisenberg uncertainty principle can be expressed as a constraint on the global symplectic
invariants $\Delta \equiv \det A + \det B + 2\det C$ (seralian)
and $\det \sigma$ \cite{sim}:
\begin{eqnarray}
\Delta\le\frac{1}{4}
+ 4\det \sigma, \label{heis}\end{eqnarray}
and the symplectic eigenvalues
$\nu_{\mp},$ which form the symplectic spectrum of covariance matrix $\sigma$
are determined by
\begin{eqnarray}
2\nu_{\mp}^2=\Delta\mp\sqrt{\Delta^2-4\det\sigma}.
\label{eigs}\end{eqnarray}
In terms of $\nu_{\mp}$ relation (\ref{heis}) takes the form $\nu_{-}\ge1/2$.
A two-mode Gaussian state
is pure if and only if $\det \sigma=1/16$. Equivalently, the necessary and sufficient criterion for a state to be pure is
$\nu_{-}=\nu_{+}=1/2.$ We notice also the useful relations
\bea
\det \sigma=
\nu_{-}^{2}\nu_{+}^{2},~~~
\Delta = \nu_{-}^{2} + \nu_{+}^{2}.
\eea

For a Gaussian density operator, the negativity is completely defined by the symplectic spectrum of the partial transpose of the covariance matrix and it is given by
\begin{eqnarray}
E_{N}={\rm max}\{0,-\log_2 2\tilde\nu_-\},
\end{eqnarray}
where $\tilde\nu_-$ is the smallest of the two symplectic eigenvalues of the partial transpose $\tilde{{\sigma}}$ of the two-mode covariance matrix $\sigma:$
\begin{eqnarray}
    2\tilde{\nu}_{\mp}^2 = \tilde{\Delta}\mp\sqrt{\tilde{\Delta}^2
-4\det\sigma}.
\end{eqnarray}
Here $ \tilde\Delta$ is the symplectic invariant, given by
$ \tilde\Delta=\Delta - 4 \det C = \det A+\det B-2\det C.$

Logarithmic negativity determines the strength of entanglement for $E_N(t)>0.$ A state is separable
if and only if \begin{eqnarray}\tilde\nu_{-}\ge1/2\label{seigen}\end{eqnarray} and logarithmic negativity quantifies the violation of inequality (\ref{seigen}).
As expected, the time evolution of the logarithmic negativity has a behaviour similar to that one of the Simon function in what concerns the characteristics of the state of being separable or entangled \cite{arus,aijqi,ascri,aosid}.

\subsection{Asymptotic entanglement}

From Eqs. (\ref{covarinf}) and (\ref{envcoe}) we obtain the following elements of the asymptotic matrices $A(\infty)=B(\infty):$
\begin{eqnarray} m\omega\sigma_{xx}(\infty)=\frac{\sigma_{p_xp_x}(\infty)}{m\omega}=\frac{1}{2}\coth\frac{\omega}{2kT}, ~~~\sigma_{xp_x}(\infty)=0
\label{varinf} \end{eqnarray}
and of the entanglement matrix $C(\infty):$
\begin{eqnarray}\sigma_{xy} (\infty) =
\frac{m^2(\lambda^2+\omega^2)D_{xy}+m\lambda
D_{xp_y}}{m^2\lambda(\lambda^2+\omega^2)},\end{eqnarray}
\begin{eqnarray}\sigma_{xp_y}(\infty)=
\sigma_{yp_x}(\infty)=\frac{\lambda
D_{xp_y}}{\lambda^2+\omega^2},\end{eqnarray}
\begin{eqnarray}\sigma_{p_xp_y} (\infty) =
\frac{m^2\omega^2(\lambda^2+\omega^2)D_{xy}-m\omega^2\lambda D_{xp_y}}{\lambda(\lambda^2+\omega^2)}.\end{eqnarray}

The mixedness of a quantum state $\rho$ is characterized by its
purity $\mu\equiv {\rm Tr}\rho^2$. For a two-mode
Gaussian state
the purity is given by
$\mu=1/4\sqrt{\det\sigma}$. The marginal purities $\mu_i$  ($i=1,2$) of the reduced states in mode $i$ are given by
$\mu_1=1/2\sqrt{\det A}$ and $\mu_2=1/2\sqrt{\det B}.$ The global and marginal purities range from $0$ to $1$,
and they fulfill the constraint $\mu \ge \mu_1 \mu_2,$ as a direct consequence of Heisenberg uncertainty relations. In Ref. \cite{ade1} the following
upper and lower bounds on the invariant $\Delta$ have been obtained in terms of global and marginal purities:
\begin{eqnarray}
\frac{1}{2 \mu} + \frac{(\mu_1 - \mu_2)^2}{4\mu_1^2 \mu_2^2}\le\Delta
\le \min \left\{ \frac{(\mu_1 + \mu_2)^2}{4 \mu_1^2 \mu_2^2}
- \frac{1}{2 \mu}, \frac{1}{4} \left(1+\frac{1}{\mu^2}\right)  \right\}.\label{ineqd}
\end{eqnarray}
The invariant $\Delta$ has a direct physical
interpretation \cite{ade1}: at given global and marginal
purities, $\Delta$ determines the amount of entanglement
of the state and the smallest symplectic eigenvalue $\tilde\nu_-$ of the partially transposed state
is strictly monotone in $\Delta.$ Consequently, the entanglement of a
Gaussian state with fixed global purity $\mu$ and marginal
purities $\mu_1,\mu_2$ is strictly increasing with decreasing $\Delta$.
According to double inequality (\ref{ineqd}), giving lower and upper
bounds on $\Delta,$ there exist both maximally and minimally
entangled Gaussian states.

We analyze now the existence of the entanglement in the asymptotic regime in the considered symmetric situation $A=B$ and in the particular case $D_{xy}=0.$ In the limit of long times, we obtain:
\begin{eqnarray}
\det A = \det B = \frac{C_T^2}{4},~~~\det C = - \frac {d^2}{\Lambda^2},~~~
\det\sigma = \left(\frac{C_T^2}{4} - \frac{d^2}{\Lambda^2}\right)^2,\end{eqnarray}
where we used the notations:
\begin{eqnarray}
C_T \equiv \coth\frac {\omega}{2kT},~~~ d \equiv D_{xp_y},~~~\Lambda^2 \equiv \omega^2 + \lambda^2.\end{eqnarray}
Then we get
\begin{eqnarray}
\frac{1}{\mu}=C_T^2 - 4\frac{d^2}{\Lambda^2},~~~\frac{1}{\mu_1} =\frac{1}{ \mu_2} = C_T,~~~\Delta = 2\sqrt{\det\sigma}=\frac{1}{2\mu}.
\end{eqnarray}
These values saturate the lower bound in inequalities (\ref{ineqd}) and this situation entails a maximal entanglement. Consequently, the corresponding states are Gaussian maximally entangled mixed states (GMEMS). They are thermal squeezed states with the squeezing parameter $r$ given by $\tanh 2r = d/\Lambda C_T. $ In the pure case ($\det \sigma = 1/16$), these states are equivalent to two-mode squeezed vacua with the squeezing parameter determined only by the temperature of the environment :
$\tanh 2r = \sqrt{C_T^2-1}/2C_T. $

According to Ref. \cite{ade1},  these states are separable in the range
\begin{equation}
\mu \le \frac{\mu_1 \mu_2}{\mu_1 + \mu_2 - \mu_1 \mu_2}.\label{sepreg}
\end{equation}
Then, for a given temperature $T,$ we obtain that the asymptotic final state is separable for the following range of positive values of the mixed diffusion coefficient $d$:
\begin{eqnarray}
\frac{2d}{\Lambda}\le C_T-1.\label{sep}\end{eqnarray} We remind that, according to inequalities (\ref{coefineq}), the coefficients have to fulfill also the constraint $\lambda C_T/2 \ge d.$
For a given temperature of the environment and for this range of mixed diffusion coefficients,
no entanglement can occur for these states. Outside this region, i.e. for a temperature and diffusion coefficient
satisfying \begin{eqnarray}
C_T-1\le\frac{2d}{\Lambda}\le C_T+1,\label{entan}\end{eqnarray}
they are GMEMS.

The asymptotic logarithmic negativity has the form
\begin{eqnarray} E_N(\infty)=-\log_2\left(C_T-\frac{2d}{\Lambda}\right).\end{eqnarray}
Outside the separable region, this is the maximum possible value of the
logarithmic negativity, attained by GMEMS.
It depends only on the mixed diffusion coefficient, dissipation constant and temperature, and does not depend on the parameters of the initial Gaussian state.

\subsection{Entropy and mutual information}

The degree of mixedness of a quantum state
$\rho$ can be characterized
either by von Neumann entropy
$S_{V}({\rho})$ or by
the linear entropy $S_{L}({\rho})$.
For CV systems these quantities are defined as
\begin{eqnarray}
S_{V}({\rho})
\equiv - {\rm Tr}{\rho}
\ln{\rho},~~~
S_{L}({\rho})  \equiv  1 -
{\rm Tr}{\rho}^{2}
\equiv 1 - \mu.
\label{linea}
\end{eqnarray}

In Ref. \cite{ser2} it was proven that von Neumann entropy
$S_{V}({\sigma})$ of an arbitrary two-mode Gaussian state with covariance matrix
$\sigma$ has the following expression:
\begin{equation}
S_{V}(\sigma)=f(\nu_{-})+
f(\nu_{+}),
\label{magicformula}
\end{equation}
with $\nu_{\mp}$ given
by Eqs. (\ref{eigs}) and
\bea
f(x)\equiv(x+\frac12)\ln(x +\frac12)
-(x-\frac12)\ln(x -\frac12).
\label{magicfunc}
\eea

In the symmetric case considered in the previous Subsection, corresponding to a maximal entanglement, we obtain the following expression of the von Neumann entropy of GMEMS:
\bea
S_{V}^M(\sigma)=(2\nu_M + 1)\ln(\nu_M +\frac12)
-(2\nu_M - 1)\ln(\nu_M - \frac12),
\label{entro}
\eea
where \bea \nu_M\equiv \nu_-=\nu_+= \sqrt{\frac{C_T^2}{4}-\frac{d^2}{\Lambda^2}}.\eea
We see that von Neumann entropy
of a two-mode Gaussian state
depends on the two invariants
$\Delta$ and
$\det \sigma$,
whereas the purity depends only on $\det \sigma.$ This implies that the hierarchy of mixedness
established by von Neumann entropy on the set of
two-mode Gaussian states differs from that induced by the
linear entropy. Namely, there may exist states with a
given linear entropy, i.e. with a given
$\det\sigma$, but with different
von Neumann entropies, i.e. with different
$\Delta$. Therefore von Neumann entropy
provides a richer characterization of the lack of information of a state \cite{ser2}.

Knowing the von Neumann
entropy one can
obtain the mutual information
$I(\rho) \equiv
S_{V}(\rho_{1})
+ S_{V} (\rho_{2})
- S_{V}(\rho)$
(here $\rho_{i}$
is the reduced density matrix of
subsystem $i=1,2$), which
quantifies the total amount of correlations
(both quantum and classical) contained in a
state \cite{ser2}. The mutual information $I(\sigma)$ of
a Gaussian state with covariance matrix $\sigma$ is defined as
\begin{equation}
I(\sigma) = S_{V}(\sigma_{1})+
S_{V}(\sigma_{2}) - S_{V}(\sigma),
\label{defmi}
\end{equation}
where $\sigma_{i}$
denotes the reduced one-mode state obtained
by tracing over subsystem $j\neq i$.
For an arbitrary two-mode
Gaussian state the mutual information
$I(\sigma)$ is given by \cite{ser2}:
\begin{equation}
I(\sigma)=f(a)+f(b)-f(\nu_{-})
-f(\nu_{+}),
\label{misf}
\end{equation}
where $a=\sqrt{{\det}~A}$,
$b=\sqrt{{\det}~B},$ and
$f(x)$ is given by Eq. (\ref{magicfunc}).

For a symmetric state with covariance matrix $\sigma_{s}$,
i.e. a state with $a=b$, and for the case corresponding to a maximal entanglement, we obtain for the mutual information of GMEMS the following expression:
\bea
I^M(\sigma_{s})=(2a + 1)\ln(a +\frac12)
-(2a - 1)\ln(a - \frac12) - (2\nu_M + 1)\ln(\nu_M +\frac12)
+ (2\nu_M - 1)\ln(\nu_M - \frac12),
\label{mutu}
\eea
where $a = C_T/2.$

\section{Summary}
\pst

In the framework of the theory of open quantum systems based on completely positive quantum dynamical semigroups, we investigated the Markovian dynamics of the quantum entanglement for a subsystem composed of two noninteracting modes embedded in a thermal environment. By using the Peres-Simon necessary and sufficient criterion of separability for two-mode Gaussian states, we have described the entanglement in terms of the covariance matrix for Gaussian input states.
For a given temperature, we calculated the range of mixed diffusion coefficients which determine the existence of asymptotic Gaussian maximally entangled mixed states (GMEMS). We obtained also the expression of the maximal logarithmic negativity, which characterizes the degree of entanglement of these states.
Using the symplectic
eigenvalues of the asymptotic covariance matrix, we have characterized mixedness and total amount of correlations (both quantum and classical) contained in asymptotic GMEMS by determining von Neumann entropy and mutual information for these states.

\section*{Acknowledgments}
\pst

The author acknowledges the financial support received within
the Project CNCSIS-IDEI 497/2009 and Project PN 09 37 01 02/2009.

\end{document}